\documentstyle{article}
\title{{\protect\null\hfill \large TSPI-TH1/94} \\
\protect\vspace{1cm}
{\LARGE \bf Massive fields dynamics in open bosonic
string theory\\ \protect\vfill} \author{I.L.BUCHBINDER \\ Department of
Theoretical Physics, \\ Tomsk State Pedagogical Institute, Tomsk
634041, Russia\\{}\\ V.A.KRYKHTIN and V.D.PERSHIN\\Department of
Quantum Field Theory,\\ Tomsk State University, Tomsk 634050, Russia}}
\date{}

\begin{document}
\maketitle
\vfill
\begin{abstract}
We consider the theory of open bosonic string in massive background fields.
The general structure of renormalization is investigated. A general
covariant action for a string in background fields of the first massive
level is suggested and its symmetries are described. Equations of motion
for the background fields are obtained by demanding that the renormalized
operator of the energy-momentum tensor trace vanishes.
\end{abstract}
\vfill
\thispagestyle{empty}

\newpage
\setcounter{page}{1}

\section{\bf Introduction}
The theory of a string in massless background fields provides a possibility
to consider string interactions in the low energy approximation [1-4].
Such a theory is remarkable due to its close connection with the
two-dimentional quantum field theory and possible generalizations (see
review \cite{5}). Namely, this approach has led to the equations of
string gravity which are now widely used for finding new cosmological
solutions (see review \cite{6}).

The crucial point of the string theory in massless background fields is the
fundamental concept of Weyl invariance. In quantum theory it means that the
renormalized operator of the energy-momentum tensor trace must vanish
resulting in equations of motion for background fields. The general
analysis of the renormalized operator for a bosonic string interacting
with background metric, antisymmetric tensor and dilaton was performed
in refs.\cite{7,8} (see also review \cite{5}).

A natural development of the approach leads to the consideration of
string in background fields which are connected with massive modes in
the string spectrum. Unfortunately, construction of a consistent
quantum theory in this case appears to be quite difficult. A string
interacting with any finite number of massive background fields is
non-renormalizable theory and requires infinite number of
countertermes. So we have to deal with the theory containing infinite
number of terms in classical action which describe interaction with
background fields of all the massive modes. The only massive field that
does not require infinite number of counterterms is the field of
tachyon but in this case non-perturbative effects play a crucial role
[9-13] (see also review \cite{14}).

Recently several attempts were undertaken to describe string in massive
background fields [12-23]. All these investigations (excluding work on
the tachyon problem) concerned mainly only linear equations of motion for
background fields. The linear approximation is of great importance
since the equations for background fields in this approximation should
correspond to the known equations defining the string spectrum. It
was turned out that even at the linear level the whole clarity is
absent and the very possibility of going beyond the linear
approximation presents difficulties.

In ref.\cite{24} we proposed an approach to the string theory in massive
background fields that represents a direct generalization of the
$\sigma$-model approach to the string theory in massless background fields.
For a closed string interacting with background fields of all the massive
modes we showed that the renormalization has a special structure. Namely,
the renormalization of background fields of any massive level requires
consideration only of background fields of this level and of all the lower
ones, but is not affected by the infinite number of background fields
belonging to higher levels. In principle, our approach allows to go beyond
the linear approximation.

In the same paper \cite{24} we examined in detail a closed bosonic string
in background fields of the first massive level and received linear
equations of motion, though the problem of agreement with the string
spectrum was not solved completly. It is notable that the lagrangian of a
closed bosonic string in massive background fields constructed in
ref.\cite{24} appeared then to be useful for the formulation of a
generalized model of two-dimentional dilaton gravity \cite{25}.

This paper is devoted to further development of our approach \cite{24} and
to its application to the theory of open string in massive background
fields.  The theory of an open string is a field theory in space-time
with boundary.  Various aspects of quantum calculations in the theory
of open strings were discussed in ref. \cite{29}. As was noted in
pioneer works \cite{2}, interaction of an open string with background
fields corresponding to the open string spectrum is completely
concentrated at the boundary of the world sheet. Detailed investigation
of open string in massless background fields was conducted in
refs.[26-28]. Questions of quantum field theory on a manifold with
boundary were studied in recent works \cite{30,31}.

The paper is organized as follows. Section 2 deals with the investigation
of renormalization in the theory of a string interacting with arbitrary
background fields both on the world sheet and its boundary. It is showed
that the renormalization has the same structure as in ref.\cite{24}. In
section 3 we consider an open string in background fields of the first
massive level, introduce the most general action, discuss its symmetries
and conduct the renormalization of background fields. The renormalization
of composite operators defining the energy-momentum tensor trace is carried
out and linear equations of motion for background fields are derived in
Section 4. Completele agreement with the equations specifying the string
spectrum is established.

\section{\bf General analysis of renormalization}
Our aim consists in construction of a $\sigma$-model type action describing
interaction of an open string with massive background fields and deriving
from the quantum Weyl invariance condition effective equations of motion
for these fields. So we have to build up the renormalized operator of the
energy-momentum tensor trace and to demand that it vanishes.

As shown in ref.\cite{24}, to make the theory renormalizable in the case of
a closed string one has to consider an action comprising infinite set
of terms describing interaction with all possible background fields. After
rescaling string coordinates $x^{\mu}\to\sqrt{\alpha'}x^{\mu}$ the total
action takes the form
\begin{eqnarray}
\label{1}
S&=&\int\limits_{M^2} \!\!\! d^2\! z  \sqrt{g} \sum_{n=0}^{\infty} (\alpha ')^n
\sum_{i_n=1}^{N_n} {\cal O}_{i_n}^{(n)}(z,\partial x(z)) B_{i_n}^{(n)}(x(z))
\end{eqnarray}
$B_{i_n}^{(n)}$ are background fields corresponding to the $n$-th level in
the closed string spectrum, ${\cal O}_{i_n}^{(n)}$ are constructed from
$g^{ab}, \epsilon^{ab}, \partial x, D\partial x, \ldots, R^{(2)}, \partial
R^{(2)}, D\partial R^{(2)}$, \ldots . Here the derivative $D_a$ is
covariant under reparametrizations both on the world sheet and in the
D-dimensional space-time:
\begin{eqnarray}
D_a\partial_bx^\mu=\partial_a\partial_bx^\mu-\Gamma^c_{ab}(g)
\partial_cx^\mu+\mbox{\boldmath$\Gamma$}^\mu_{\lambda\rho}(G)
\partial_ax^\lambda\partial_bx^\rho .
\end{eqnarray}
For each $n$ dimension of all ${\cal O}_{i_n}^{(n)}$ in two-dimensional
derivatives equals $2n+2$. $N_n$ is the total number of all independent
${\cal O}_{i_n}^{(n)}$ belonging to the $n$-th level. The integral in
(\ref{1}) is taken over the whole world sheet, and we use a euclidian
metrics.

In case of an open string there appears a possibility to introduce
interaction at the boundary of the world sheet $\partial M$ and the total
action should be of the form
\begin{eqnarray}
\nonumber
S&=&\int\limits_{M^2} \!\!\! d^2\! z  \sqrt{g} \sum_{n=0}^{\infty} (\alpha ')^n
\sum_{i_n=1}^{N_n} {\cal O}_{i_n}^{(n)}(z,\partial x(z))
B_{i_n}^{(n)}(x(z)) +
\\
\label{3}
& &+\int\limits_{\partial M} \!\!\! dt e \sum_{k=0}^{\infty}(\alpha')^{k/2}
\sum_{i_k=1}^{N_{k}^{B}} {\cal O}_{i_k}^{B(k)}(t, x(t))
B_{i_k}^{B(k)}(x(t))
\end{eqnarray}
Here $B_{i_k}^{B(k)}$ are background fields belonging to the $n$-th massive
level of the open string, ${\cal O}_{i_k}^{B(k)}$ are constructed from
$\dot{x}^\mu = \frac{dx^\mu}{edt}$, $\ddot{x}^\mu$, \ldots, the external
curvature of the boundary $K(t)=e^{-2}n_a\left(\frac{d^2z^a}{dt^2}+
\Gamma^a_{bc}\frac{dz^b}{dt}\frac{dz^c}{dt}\right)$ and its derivatives.
$t$ is a parametr along the boudary, $n_a$ is a unit vector normal to
it and $e^2(t)=g_{ab}(z(t))\frac{dz^a}{dt}\frac{dz^b}{dt}$ is
one-dimensional metrics. ${\cal O}_{i_k}^{B(k)}$ may contain any powers
of derivatives with respect to $t$, that is why the second integral in
(\ref{3}) is expanded in powers of $\alpha'^{1/2}$.

To construct the renormalized operator of the energy-momentum tensor trace
one has to renormalize both the background fields $B_{i_k}$ and the
composite operators ${\cal O}_{i_k}$. Renormalization of fields is
constructed by demanding that divergences of the quantum effective action
vanish. In one-loop approximation it appears as
\begin{eqnarray}
\label{4}
\Gamma^{(1)}=S +\frac{1}{2} Tr\,ln \|{\cal H}_{\mu\nu}\| ,
\end{eqnarray}
where $S$ is the classical action and ${\cal H}_{\mu\nu}$ represents its
second functional derivative:
\begin{eqnarray}
{\cal H}_{\mu\nu}=\frac{\delta^2S}{\delta x^\mu \delta x^\nu} .
\end{eqnarray}
In our case (\ref{3}) ${\cal H}_{\mu\nu}$ is an operator of the following
form:
\begin{eqnarray}
\nonumber
{\cal H} \sim -D^2 &+& \sum_{k=0}^{\infty} P^{a_1\ldots a_{2k}} D_{a_1}
\ldots D_{a_{2k}} +
\\&+&\sum_{k=0}^{\infty} V_k  (\frac{dz^b}{edt}D_b)^k
\\
\nonumber
& &D^2=g^{ab} D_a D_b
\end{eqnarray}
Coefficients $P^{a_1\ldots a_{2k}}$ and $V_k$ are functions of background
fields and can be presented as series in powers of $\alpha'$:
\begin{eqnarray}
P=\sum_{n=0}^{\infty} (\alpha ')^n P^{(n)}, \hspace{5em}
V=\sum_{m=0}^{\infty} (\alpha ')^{m/2} V^{(m)},
\end{eqnarray}
where each term of expansions depends only on fields of the given massive
level.

Divergences in (\ref{4}) appear from the expression
\begin{eqnarray}
\nonumber
Tr\,ln H &\sim& Tr\,ln(-D^2)
\\ &-& \sum_{l=1}^{\infty} \frac{1}{l} Tr \left(
\nonumber
\sum_{k=0}^{\infty} P^{a_1\ldots a_{2k}} D_{a_1}\ldots
D_{a_{2k}} \frac{1}{D^2}+\right.
\\
&+&\left. \sum_{k=0}^{\infty} V_k
\left(\frac{dz^b}{edt}D_b\right)^k \frac{1}{D^2} \right)^l
\end{eqnarray}
Being local constructions, the divergences must be expanded in the same set
of ${\cal O}_{i_n}$ as the action (\ref{3})
\begin{eqnarray}
\nonumber
& &\sum_{n=0}^{\infty} (\alpha ')^n\int\limits_{M^2}\!\!\! d^2\!
z\sqrt{g} \sum_{i_n=1}^{N_n} {\cal O}_{i_n}^{(n)}{\cal
T}_{i_n}^{(n)}(B)+
\\
& &+\sum_{k=0}^{\infty}\alpha
'^{k/2}\int\limits_{\partial M}\!\!\! dte \sum_{i_k=1}^{N_{k}^{B}}
{\cal O}_{i_k}^{B(k)}{\cal T}_{i_k}^{B(k)}(B), \end{eqnarray} where
${\cal T}_{i_n}^{(n)}(B)$, ${\cal T}_{i_k}^{B(k)}(B)$, are some
dimensionless functions of background fields.

It is obvious from dimensional considerations that counterterms of some
given power in $\alpha'$ can depend only on background fields of the
corresponding massive level and of all the lower ones. Therefore to
renormalize background fields of the $n$-th massive level of the closed
string it is sufficient to calculate divergences generated by background
fields from the $n$-th and all the lower levels of the closed string
spectrum and by fields from the $2n$-th and all the lower levels of the
open string spectrum. Similarly, renormalization of $k$-th level fields of
the open string requires to consider the open string spectrum fields of the
$k$-th and all the lower levels and the closed string spectrum fields of
the [$k/2$]-th and all the lower levels ([~] means an integer part of a
number).

For example, to renormalize the fields of the first massive open string
level one is to study divergences generated by these fields and by the
massless ones. Moreover, if we are interested merely in linear
approximation it is suffitient to restrict ourselves to the study of
divergences generated by fields of the only given massive level.
Contributions of all other levels contain products of several background
fields and so are beyond the linear approximation.

\section{\bf Action, symmetries, renormalization}
In this section we will consider an open string interacting with all the
massive fields of the first level, describe its symmetries and construct
one-loop renormalization of background fields. The action should be the sum
of the free string action $S_0$ and the action $S_I$ describing interaction
with the fields of the first massive level and containing all possible
terms of the second order in derivatives:
\begin{eqnarray}
\nonumber
S[x]&=&S_0[x]+S_I[x]
\\
\nonumber
S_0[x]&=&\frac{1}{4\pi}\int\limits_{M^2} \!\!\! d^2 \! z \sqrt{g} g^{ab}
\partial_a x^\mu \partial_b x^\nu \delta_{\mu\nu}
\\
\nonumber
S_I[x]&=&\frac{\alpha '^{1/2}}{2\pi} \int\limits_{\partial M} \!\!\! dt e(t)
\left[ A_{\mu\nu}(x)\dot{x}^\mu \dot{x}^\nu + B_{\mu}(x)\ddot{x}^\mu +
K^2\varphi_1(x) +\right.
\\
\label{3.1}
& &+\left.\dot{K}\varphi_2(x) +K\dot{x}^\mu \varphi_{\mu}(x)\right]
\end{eqnarray}
Massless fields do not contribute to renormalization of massive fields in
the linear approximation and so are not considered here. A theory similar
to (\ref{3.1}) was investigated in \cite {16}, but it did not contain all
possible background fields and calculations were not explicitly covariant.

Possibility to add an arbitrary total derivative to the lagrangian in
(\ref{3.1}) yields to the symmetry of the theory under the following
transformations:
\begin{eqnarray}
\label{3.2}
\left\{
\begin{array}{l}
\delta A_{\mu\nu} =\partial_{(\mu}\Lambda_{\nu )} \\
\delta B_\mu = \Lambda_\mu
\end{array}
\hspace{5em}
\right\{
\begin{array}{l}
\delta\varphi_\mu = \partial_\mu\Lambda \\
\delta\varphi_2 = \Lambda
\end{array}
\end{eqnarray}
where $\Lambda_\mu(x)$ and $\Lambda(x)$ are arbitrary functions playing the
role of transformation parameters. As one can see, the field $B_\mu(x)$ and
$\varphi_2(x)$ are Stuckelberg ones and the symmetry allows to choose them
to be equal to zero. Thus the essential background fields are
$A_{\mu\nu}(x), \varphi_\mu(x), \varphi_1(x)$.

To renormalize background fields it is necessary to calculate divergences
of the action
\begin{eqnarray}
\label{3.3}
\frac{1}{2}Tr\,ln(S_{0\alpha\beta}+\frac{\alpha'^{1/2}}{2\pi}V_{\alpha\beta}),
\end{eqnarray}
where we denote
\begin{eqnarray}
S_{0\alpha\beta}\equiv\frac{\delta^2S_0[x]}{\delta x^\alpha\delta x^\beta},
\hspace{3em}
\frac{\alpha'^{1/2}}{2\pi}V_{\alpha\beta}\equiv
\frac{\delta^2S_I[x]}{\delta x^\alpha\delta x^\beta}
\end{eqnarray}
(\ref{3.3}) is expanded into series in powers of $(\alpha')^{1/2}$:
\begin{eqnarray}
\nonumber
\frac{1}{2}Tr\,ln(S_{0\alpha\beta}+\frac{\alpha'^{1/2}}{2\pi}V_{\alpha\beta})=
\\
\label{3.5}
=\frac{1}{2}Tr\,lnS_{0\alpha\beta}+\frac{\alpha'^{1/2}}{2}TrV_{\alpha\gamma}
G^{\alpha\beta}+O(\alpha'),
\end{eqnarray}
where Green function $G^{\alpha\beta}$ of a free string is determined by
the equation
\begin{eqnarray}
2\pi S_{0\alpha\beta}G^{\beta\gamma}=\delta^\gamma_\alpha .
\end{eqnarray}
Divergences of the first term in (\ref{3.5}) are cancelled by the
corresponding contribution of the ghosts providing that D$=26$. Terms
$O(\alpha')$ contribute to renormalization of background fields of the
second and all the higher levels and so will be omitted.

Choose coordinates $(t,y)$ on the world sheet so that $t$ be a parameter
along the boundary $\partial M$ and $y$ equal the distance between the
point $z^a=(t,y)$ and the boundary along a geodesic line targent to the
internal normal vector $n_a$. The metrics in terms of these coordinates is
\begin{eqnarray}
ds^2=e^2(t,y)dt^2+dy^2.
\end{eqnarray}
Specifying the coordinate $t$ so that $e(t,y)|_{y=0}=1$ we get the
following expansion \cite{30}:
\begin{eqnarray}
e(t,y)=1-K(t)y-\frac{1}{4}R(t,0)y^2+O(y^3)
\end{eqnarray}
Calculation of $V_{\alpha\beta}$ in such coordinates gives
\begin{eqnarray}
\label{3.9}
V_{\alpha\beta}(t,y;t',y')=\delta_{\partial M}(z)
\delta_{\partial M}(z')\sum_{k=0}^2V_{(k)\alpha\beta}
\frac{d^k\delta(t-t')}{dt^k}
\end{eqnarray}
where
\begin{eqnarray}
\nonumber
V_{(2)\alpha\beta}&=&B_{\alpha,\beta}+B_{\beta,\alpha}-2A_{\alpha\beta}
\\
\nonumber
V_{(1)\alpha\beta}&=&K\Phi_{[\alpha\beta]}+2C_{\alpha(\beta\mu)}\dot{x}^\mu
\\
\nonumber
V_{(0)\alpha\beta}&=&K^2\varphi_{1,\alpha\beta}+\dot{K}\Phi_{\alpha,\beta}
+K\Phi_{[\alpha\mu],\beta}\dot{x}^\mu+
\\
& &+C_{\alpha(\mu\nu),\beta}\dot{x}^\mu\dot{x}^\nu
+H_{(\mu\alpha),\beta}\ddot{x}^\mu
\end{eqnarray}
and
\begin{eqnarray}
\nonumber
H_{\alpha\beta}&\equiv&B_{\alpha,\beta}+B_{\beta,\alpha}-2A_{\alpha\beta},
\\
\nonumber
C_{\mu(\alpha\beta)}&\equiv&A_{\alpha\beta,\mu}-A_{\mu\alpha,\beta}
-A_{\mu\beta,\alpha}+B_{\mu,\alpha\beta}
\\
\nonumber
\Phi_{[\mu\alpha]}&\equiv&\varphi_{\alpha,\mu}-\varphi_{\mu,\alpha}
\\
\Phi_\mu&\equiv&\varphi_{2,\mu}-\varphi_\mu.
\end{eqnarray}
Here the function $\delta_{\partial M}(z)$ is defined as follows
\begin{eqnarray}
\frac{\delta x^\mu(t)}{\delta x^\nu(z')}=\delta(t-t')
\delta_{\partial M}(z')\delta^\mu_\nu .
\end{eqnarray}
Considering (\ref{3.9}), the divergences of (\ref{3.5}) are contained in
the expression
\begin{eqnarray}
\nonumber
\left. \frac{\alpha'^{1/2}}{2} \sum_{k=0}^2 \int\!\!\!dtV_{(k)\alpha\beta}(t)
\frac{d^k}{dt^k} G^{\beta\alpha}(t,0;t',0) \right|_{t'\to t}=
\\
\label{3.12}
=-\frac{\mu^\epsilon}{\epsilon}\frac{\alpha'^{1/2}}{2\pi}
\int dt V_{(0)\alpha\beta}(t)\eta^{\alpha\beta},
\end{eqnarray}
where we used that the divergences of Green function in coincident points
in the framework of dimensional renormalization are \cite{30,31}:
\begin{eqnarray}
\nonumber
\left. G^{\mu\nu}(t,0;t',0) \right|^{div}_{t'\to t} &=&
- \frac{\mu^\epsilon}{\pi\epsilon} \eta^{\mu\nu} ,
\\
\label{3.13}
\frac{d}{dt} \left. G^{\mu\nu}(t,0;t',0) \right|^{div}_{t'\to t}&=&
\frac{d^2}{dt^2} \left. G^{\mu\nu}(t,0;t',0) \right|^{div}_{t'\to t}=0.
\end{eqnarray}
Omitting in (\ref{3.12}) total derivatives we arrive at the following
one-loop effective action:
\begin{eqnarray}
\nonumber
\Gamma^{(1)}&=&\frac{\alpha'^{1/2}\mu^\epsilon}{2\pi}\int\limits_{\partial M}
\!\!\!dte\left(\dot{x}^\mu\dot{x}^\nu(\stackrel{\circ}{A}_{\mu\nu}
-\frac{1}{\epsilon}\Box\stackrel{\circ}{A}_{\mu\nu})+\right.
\\
\label{3.14}
&+&\ddot{x}^\mu(\stackrel{\circ}{B}_\mu
-\frac{1}{\epsilon}\Box\stackrel{\circ}{B}_\mu)
+K^2(\stackrel{\circ}{\varphi}_1
-\frac{1}{\epsilon}\Box\stackrel{\circ}{\varphi}_1)+
\\
\nonumber
&+&\left.\dot{K}(\stackrel{\circ}{\varphi}_2
-\frac{1}{\epsilon}\Box\stackrel{\circ}{\varphi}_2)
+K\dot{x}^\alpha(\stackrel{\circ}{\varphi}_\alpha
-\frac{1}{\epsilon}\Box\stackrel{\circ}{\varphi}_\alpha)\right)+(fin),
\\
\nonumber
& &\Box\equiv\eta^{\alpha\beta}\partial_\alpha\partial_\beta,
\end{eqnarray}
where $\circ$ denotes bare background fields and $(fin)$ stands for a
finite part of the one-loop correction.

To cancel the divergences in (\ref{3.14}) renormalization of all the
background fields should be of the form:
\begin{eqnarray}
\label{3.15}
\stackrel{\circ}{\Phi} = \mu^{-\epsilon}(\Phi + \frac{1}{\epsilon} \Box \Phi )
\end{eqnarray}
where $\Phi=(A_{\mu\nu},B_\mu,\varphi_1,\varphi_2,\varphi_\mu)$.

\section{\bf Renormalization trace of energy-momentum tensor and equations
of motions}
In classical theory trace of the energy-momentum tensor for the theory
(\ref{3.1}) on the $2+\epsilon$-dimensional world sheet is
\begin{eqnarray}
\nonumber
T(z) &=& g_{ab}(z) \frac{\delta S}{\delta g_{ab}(z)} =
\\
\nonumber
&=&\frac{\epsilon}{8\pi} g^{ab}(z) \partial_a x^\mu \partial_b x^\nu
\eta_{\mu\nu}
+ \frac{\alpha '^{1/2}}{8\pi} H_{\mu\nu} \dot{x}^\mu \dot{x}^\nu
\delta_{\partial M}(z)-
\\
\nonumber
&-& \frac{\alpha '^{1/2}}{4\pi} K^2 \varphi_1 \delta_{\partial M}(z)
+ \frac{\alpha '^{1/2}}{2\pi} K \varphi_1 \delta_{\partial M}'(z)+
\\
&+& \frac{\alpha '^{1/2}}{4\pi} K \dot{x}^\mu \Phi_\mu \delta_{\partial M}(z)
- \frac{\alpha '^{1/2}}{4\pi} \dot{x}^\mu \Phi_\mu \delta_{\partial M}'(z)
+ O(\epsilon \alpha'^{1/2}).
\end{eqnarray}
Terms $O(\epsilon\alpha'^{1/2})$ will not contribute to the renormalized
trace of energy-mo\-men\-tum tensor .

To calculate the trace of the energy-momentum tensor in quantum theory we
should define renormalized values for composite operators. Consider, for
example, the vacuum average for one of these operators:
\begin{eqnarray}
\nonumber
<H_{\mu\nu}(x)\dot{x}^\mu\dot{x}^\nu>=\int Dxe^{-S[x]}H_{\mu\nu}(x)
\dot{x}^\mu\dot{x}^\nu
\biggm/
\int Dxe^{-S[x]}.
\end{eqnarray}
Making the shift $x^\mu=\bar{x}^\mu+\zeta^\mu$ in the functional integral
($\bar{x}^\mu$ are solutions of the classical equations of motion) and
using (\ref{3.13}) we get in linear approximation:
\begin{eqnarray}
<H_{\mu\nu}(x)\dot{x}^\mu\dot{x}^\nu>=\mu^\epsilon\dot{\bar{x}}^\mu
\dot{\bar{x}}^\nu(H_{\mu\nu}(\bar{x})-\frac{1}{\epsilon}\Box
H_{\mu\nu}(\bar{x}))+(fin).
\end{eqnarray}
Renormalized operators should have a finite average value
\begin{eqnarray}
<[H_{\mu\nu}(x)\dot{x}^\mu\dot{x}^\nu]>=H_{\mu\nu}(\bar{x})
\dot{\bar{x}}^\mu\dot{\bar{x}}^\nu+ (fin)
\end{eqnarray}
hence
\begin{eqnarray}
(H_{\mu\nu}(x)\dot{x}^\mu\dot{x}^\nu)_\circ=\mu^\epsilon[\dot{x}^\mu
\dot{x}^\nu(H_{\mu\nu}(x)-\frac{1}{\epsilon}\Box H_{\mu\nu}(x))]+(fin),
\end{eqnarray}
or, using (\ref{3.15}),
\begin{eqnarray}
(\stackrel{\!\!\!\!\!\circ}{H_{\mu\nu}} \dot{x}^\mu \dot{x}^\nu)_\circ =
[H_{\mu\nu} \dot{x}^\mu \dot{x}^\nu ]
\end{eqnarray}
In the same way one can get
\begin{eqnarray}
(\stackrel{\!\!\circ}{\varphi_1})_\circ=[ \varphi_1],\hspace{5em}
(\stackrel{\circ}{\Phi}_\mu \dot{x}^\mu)_\circ=[ \Phi_\mu \dot{x}^\mu ]
\end{eqnarray}
Similar but more tedious calculation give the following renormalization of
the operator $g^{ab}(z)\partial_ax^\mu\partial_bx^\nu\eta_{\mu\nu}$:
\begin{eqnarray}
\nonumber
(g^{ab}\partial_ax^\mu\partial_bx^\nu\eta_{\mu\nu})_\circ &=&
[g^{ab}\partial_ax^\mu\partial_bx^\nu\eta_{\mu\nu}]+
\\
\nonumber
&+&\alpha'^{1/2}\frac{\mu^\epsilon}{2\epsilon}
[-V_{(2)} \delta_{\partial M}''(z) + K V_{(2)} \delta_{\partial M}'(z)
\\
\nonumber
&-&\frac{1}{2} R V_{(2)} \delta_{\partial M}(z)-
- K^2 V_{(2)} \delta_{\partial M}(z) +
\\
\nonumber
&+&\frac{d^2}{dt^2} V_{(2)} \delta_{\partial M}(z) + 4 V_{(0)}
\delta_{\partial M}(z) ] ,
\\
& &V_{(2)}\equiv V_{(2)\mu\nu}\eta^{\mu\nu},\hspace{3em}
V_{(0)}\equiv V_{(0)\mu\nu}\eta^{\mu\nu}.
\end{eqnarray}
As a result, the renormalized operator of the energy-momentum tensor trace
has the form
\begin{eqnarray}
\nonumber
[T]&=&\alpha'^{1/2}\delta(y)[RE^{(0)}(x)+\dot{x}^\mu\dot{x}^\nu
E^{(1)}_{\mu\nu}(x)+\ddot{x}^\mu E^{(2)}_\mu(x)+
\\
\nonumber
&+&K\dot{x}^\mu E^{(3)}_\mu+K^2E^{(4)}(x)+\dot{K}E^{(5)}(x)]+
\\
\nonumber
&+&\alpha'^{1/2}\delta'(y)[\dot{x}^\mu E^{(6)}(x)+KE^{(7)}(x)]+
\\
\label{4.10}
&+&\alpha'^{1/2}\delta''(y)E^{(8)}(x)
\end{eqnarray}
with
\begin{eqnarray}
\nonumber
E^{(0)}(x)&=&-\frac{1}{8\pi}(B^\alpha_{,\alpha}-A^\alpha_\alpha)
\\
\nonumber
E^{(1)}_{\mu\nu}(x)&=&\frac{1}{4\pi}(2\Box A_{\mu\nu}-2A_{\mu\nu}
-A^\alpha_{\alpha,\mu\nu}-2A^\alpha_{\mu,\alpha\nu}
-2A^\alpha_{\nu,\alpha\mu}+
\\
\nonumber
& &+3B^\alpha_{,\mu\nu\alpha}+B_{\mu,\nu}+B_{\nu,\mu})
\\
\nonumber
E^{(2)}_\mu(x)&=&\frac{1}{4\pi}(2\Box B_\mu+3B^\alpha_{,\alpha\mu}
-A^\alpha_{\alpha,\mu}-4A^\alpha_{\mu,\alpha})
\\
\nonumber
E^{(3)}_\mu(x)&=&\frac{1}{2\pi}(\varphi_{2,\mu}-\varphi_\mu
+\Box\varphi_\mu-\varphi^\alpha_{,\mu\alpha})
\\
\nonumber
E^{(4)}(x)&=&\frac{1}{4\pi}(2\Box\varphi_1-B^\alpha_{,\alpha}
+A^\alpha_\alpha-2\varphi_1)
\\
\nonumber
E^{(5)}(x)&=&\frac{1}{2\pi}(\Box\varphi_2-\varphi^\alpha_{,\alpha})
\\
\nonumber
E^{(6)}_\mu(x)&=&-\frac{1}{2\pi}(\varphi_{2,\mu}-\varphi_\mu)
\\
\nonumber
E^{(7)}(x)&=&\frac{1}{4\pi}(B^\alpha_{,\alpha}-A^\alpha_\alpha
+4\varphi_1)
\\
\label{4.11}
E^{(8)}(x)&=&-\frac{1}{4\pi}(B^\alpha_{,\alpha}-A^\alpha_\alpha)
\end{eqnarray}
{}From the requirement of quantum Weyl invariance it follows that all the
coefficients (\ref{4.11}) at the linear independent operators in
(\ref{4.10}) should be equal to zero. Therefore equations of motion for the
background fields of the first massive level are $E(x)=0$.

Using the symmetry of the theory under the transformations (\ref{3.2}) to
make $B_\mu=0$ and $\varphi_2=0$ and returning to the dimensional string
coordinates $x^\mu\to\alpha'^{-1/2}x^\mu$ one can rewrite these equations
as
\begin{eqnarray}
\nonumber
B_\mu=0,\;\;\;\varphi_2=0,&\varphi_1=0,&\varphi_\mu=0,
\\
\Box A_{\mu\nu}-m^2A_{\mu\nu}=0,&A^\mu_\mu=0,&\partial_\mu A^\mu_\nu=0,
\end{eqnarray}
where $m^2$ is the mass of the first level.

So the equations of motion for the background fields of the first massive
level of the open string are equivalent to the equations describing a
massive field with spin 2. As is well known, the analysis of the spectrum
of open string physical states gives the same result.

\section{\bf Summary}
We have considered the general approach to the theory of an open string
interacting with massive background fields. Our analysis has shown that
there exist a consistent description of string models in background fields
corresponding to a finite number of the massive string modes. We have
proposed the most general model describing interaction of an open string
with background fields of the first massive level. The theory is invariant
under symmetry transformations of background fields.

The renormalized operator of the energy-momentum tensor trace has been
constructed. In linear approximation it has been shown that the
requirement of quantum Weyl invariance leads to equations of motion for
background fields which are consistent with the struture of the open string
spectrum at the first massive level. The approach proposed opens up
possibilities for deriving equations of motion for massive fields of higher
spins in the framework of the string theory.

\section*{\bf Acknowledgements}
I.L.B. is grateful to S.J.Gates, H.Osborn, B.A.Ovrut, J.Schnittger and
A.A.Tseytlin for useful discussions on a various aspects of the work. The
research described in this publication has been supported in parts by
International Science Foundation, grant RI1000 and Russian Foundation
for Fundamental Research, project No 94-02-03234.


\begin{thebibliography}{00}
\bibitem{1}
           C.Lovelace, Phys. Lett. B135 (1984) 75; Nucl.Phys. B273 (1986)
           413.
\bibitem{2}
           E.S.Fradkin, A.A.Tseytlin, Phys. Lett. B158 (1985) 316; Nucl.
           Phys. B261 (1985) 1.
\bibitem{3}
           C.G.Callan, D.Friedan, E.Martinec, M.J.Perry, Nucl. Phys. B262
           (1985) 593.
\bibitem{4}
           A.Sen, Phys. Rev. D32 (1985) 2102; Phys. Rev. Lett. 55 (1985)
           1846.
\bibitem{5}
           A.A.Tseytlin, Int. J. Mod. Phys. A4 (1989) 1257.
\bibitem{6}
           A.A.Tseytlin, in Proc. of Int. Workshop on String Quantum
           Gravity and Physics at the Planck Energy Scale, Erice, 1992, ed.
           N.Sanches, World Scientific, 1993.
\bibitem{7}
           A.A.Tseytlin, Phys. Lett. B178 (1986) 34; Nucl. Phys. B294
           (1987) 383.
\bibitem{8}
           H.Osborn, Nucl. Phys. B294 (1987)595; B308 (1988) 629; B363
           (1991) 486; Ann. Phys. 200 (1990) 1.
\bibitem{9}
           S.R.Das, B.Sathiapalan, Phys. Rev. Lett. 56 (1986) 2664; Phys.
           Rev. Lett. 57 (1986) 1511.
\bibitem{10}
           C.Itoi, Y.Watabiki, Phys. Lett. B198 (1987) 486
\bibitem{11}
           I.Klebanov, L.Susskind, Phys. Lett. B200 (1988) 446
\bibitem{12}
           R.Brustein, D.Nemeschansky, S.Yankielowicz, Nucl. Phys. B301
           (1988) 224
\bibitem{13}
           A.A.Tseytlin, Phys. Lett. B241 (1990) 233; B264 (1991) 311
\bibitem{14}
           A.A.Tseytlin, Int. J. Mod. Phys. A4 (1989) 4249
\bibitem{15}
           U.Ellwanger, J.Fuchs, Nucl. Phys. B312 (1989) 95
\bibitem{16}
           J.M.Labastida, M.A.H.Vozmediano, Nucl. Phys. B312 (1989) 308
\bibitem{17}
           J.Hughes, J.Liu, J.Polchinsky, Nucl. Phys. B316 (1989) 15
\bibitem{18}
           S.Jain, A.Jeviski, Phys. Lett. B220 (1989) 379
\bibitem{19}
           P.C.Argyres, C.R.Nappi, Phys. Lett. B224 (1989) 89
\bibitem{20}
           U.Ellwanger, Nucl. Phys. B322 (1990) 300
\bibitem{21}
           J.C.Lee, B.A.Ovrut, Nucl. Phys. B336 (1990) 222
\bibitem{22}
           J.C.Lee, Phys. Lett. B241 (1990) 336
\bibitem{23}
           U.Elliwanger, J.Schnittger, Int. J. Mod. Phys. A7 (1992) 3389
\bibitem{24}
           I.L.Buchbinder, E.S.Fradkin, S.L.Lyakhovich, V.D.Pershin, Phys.
           Lett. B304 (1993) 239
\bibitem{25}
           E.Elizalde, S.Naftulin, S.D.Odintsov, Phys. Lett. B323 (1994)
           124
\bibitem{26}
           E.S.Fradkin, A.A.Tseytlin, Phys. Lett. B163 (1985) 123
\bibitem{27}
           A.A.Tseytlin, Nucl. Phys. B276 (1986) 391
\bibitem{28}
           A.Abouelsaqod, C.G.Callan, C.R.Nappi, S.A.Yost, Nucl. Phys. B280
           (1987) 599
\bibitem{29}
           H.Luckock, Ann.Phys. 194 (1989) 113
\bibitem{30}
           D.M.McAvity, H.Osborn, Class, Quant. Grav. 8 (1991) 603; 8
           (1991) 1445; Nucl. Phys. B394 (1993) 728; B406 (1993) 655
\bibitem{31}
           D.M.McAvity, Class, Quant. Grav. 9 (1992) 1983
\end{thebibliography}
\end{document}